\documentclass[final,3p,times,twocolumn]{elsarticle}
 \biboptions{comma,sort&compress}
\usepackage{dcolumn}% Align table columns on decimal point
\usepackage{latexsym}
\usepackage{graphicx}
\usepackage{pstricks}
\usepackage{epsfig}
\usepackage{color}

\usepackage{color,amsmath,amssymb,exscale,epsfig}

\input epsf
\newcommand{\be}{\begin{equation}}
\newcommand{\ee}{\end{equation}}
\newcommand{\bea}{\begin{eqnarray}}
\newcommand{\eea}{\end{eqnarray}}

\newcommand{\lbl}[1]{\label{eq:#1}}
\newcommand{ \rf}[1]{(\ref{eq:#1})}
\newskip\humongous \humongous=0pt plus 1000pt minus 1000pt

\newif\ifdtup

\newcommand{\lapprox}{%
\mathrel{%
\setbox0=\hbox{$<$}
\raise0.6ex\copy0\kern-\wd0
\lower0.65ex\hbox{$\sim$}
}}
\newcommand{\gapprox}{%
\mathrel{%
\setbox0=\hbox{$>$}
\raise0.6ex\copy0\kern-\wd0
\lower0.65ex\hbox{$\sim$}
}}

\def\theequation{\arabic{section}.\arabic{equation}}

\begin{document}

\begin{frontmatter}

%\preprint{CPT-P003-2013}

\title{Matching long and short distances %at order ${\cal O}(\alpha_s)$ 
in the form factors for $K\to\pi \ell^+\ell^-$} %$K^\pm (K_S) \to \pi^\pm (\pi^0) \ell^+\ell^-$}

\author[label1]{Giancarlo D'Ambrosio}
%  \cortext[cor0]{FAPESP CNPq-Brasil PhD student fellow.}
\ead{gdambros@na.infn.it}
\address[label1]{INFN-Sezione di Napoli, Via Cintia, I-80126 Napoli, {Italy} }

\author[label2]{David Greynat}
%  \cortext[cor0]{FAPESP CNPq-Brasil PhD student fellow.}
\ead{david.greynat@gmail.com}
\address[label2]{Presently without affiliation}

\author[label3]{Marc Knecht}
%  \cortext[cor0]{FAPESP CNPq-Brasil PhD student fellow.}
\ead{marc.knecht@cpt.univ-mrs.fr}
\address[label3]{Centre de Physique Th\'{e}orique UMR 7332, CNRS/Aix-Marseille Univ./Univ. du Sud Toulon-Var \\
CNRS Luminy Case 907, 13288 Marseille Cedex 9, France}

\pagestyle{myheadings}
\markright{ }
\begin{abstract}
At order ${\cal O}(\alpha G_{\rm F})$, the amplitudes for the decays $K\to\pi \ell^+\ell^-$ involve a form factor
given by the matrix element of the time-ordered product of the electromagnetic
current with the four-quark operators describing weak non-leptonic neutral-current
transitions between a kaon and a pion. The short-distance behaviour of this time-ordered 
product, when considered at order ${\cal O}(\alpha_s)$ in the perturbative expansion of QCD,
involves terms linear and quadratic in the logarithm of the Euclidean momentum
transfer squared. It is shown how one can exactly match these short-distance features
using a dispersive representation of the form factor, with an 
absorptive part given by an infinite sum of zero-width resonances following a Regge-type spectrum.
Some phenomenology-related issues are briefly discussed.
\end{abstract}

%\pacs{11.15.Pg, 12.38.Lg, 13.25.Jx, 14.40.Rt}

\end{frontmatter}

\section{Introduction}
\setcounter{equation}{0}

The amplitude for a $\Delta S = +1$
neutral-current transition $K(k) \to \pi(p) \ell^+(p_+) \ell^-(p_-)$, 
$\ell = e , \mu$, where $(K,\pi) \in \{(K^+ , \pi^+) , (K^0 , \pi^0)\}$, 
takes the form \cite{EPdR87,DAmbrosio:1998gur,DAmbrosio:2018ytt}
\bea
\hspace{-0.6cm}
{\cal A}^{K \to \pi \ell^+ \ell^-} \! (s) \!\!&=& \!\!
- e^2 \times {\bar{\rm u}} (p_-) \gamma_\rho {\rm v} (p_+) \times \frac{1}{s}
\nonumber\\
&& \hspace{-0.6cm}
\times
\bigg\{ 
i \! \int \! d^4 x \,
\langle\pi(p)  \vert T \{  j^\rho  (0) {\cal L}^{\Delta S = 1}_{\mbox{\scriptsize non-lept}} (x)  \} \vert K(k) \rangle
\nonumber\\
&& \hspace{-0.6cm}
+ \, \left( - \frac{G_{\rm F}}{\sqrt{2}} V_{us} V_{ud} \right)
\times \frac{C_{7V} (\nu)}{4 \pi \alpha} \times s
\nonumber\\
&& \hspace{-0.6cm} 
\times \, \langle\pi(p)  \vert ({\bar s} \gamma^\rho d) (0) \vert K(k) \rangle
\bigg\}
\lbl{amplitude}
\\
\hspace{-0.6cm}
&=& \!\!
- e^2 \times {\bar{\rm u}} (p_-) ({\not\! k} + {\not\!\! p}) {\rm v} (p_+)
\times \frac{W_{K\pi}(z)}{16 \pi^2 M_K^2} 
.
\nonumber
\eea
Here $s= (k-p)^2 = (p_+ + p_-)^2 = z M_K^2$ is the square of the di-lepton invariant mass, $M_K$ denotes
the kaon mass [for our present purposes, it is not necessary to distinguish between the masses of charged
and neutral kaons], $e$ is the electric charge, $G_{\rm F}$ the Fermi constant, and $V_{ud}$, $V_{us}$ are elements of the CKM matrix. 
Each matrix element occurring in the first expression has to be evaluated in QCD
with three active flavours. In particular, the electromagnetic current $j^\rho$ is made up from the
contributions of the $u$, $d$ and $s$ quarks,
\be
\hspace{-0.005cm}
j^\rho (x) = \!\sum_{q = u,d,s} \! e_q ({\bar q} \gamma^\rho q) (x),~
e_u = +\frac{2}{3} , \, e_d = e_s = - \frac{1}{3}
.
\ee
Furthermore, ${\cal L}^{\Delta S = 1}_{\mbox{\scriptsize non-lept}} (x)$
represents the effective La\-gran\-gian for $\Delta S = +1$ transitions
\cite{Gaillard:1974nj,Altarelli:1974exa,Witten:1976kx,Shifman:1975tn,Wise:1979at,Gilman:1979bc}, 
and involves the two current-current
four-quark operators, as well as the QCD penguin operators, modulated by the appropriate Wilson coefficients
\be
{\cal L}^{\Delta S = 1}_{\mbox{\scriptsize non-lept}} (x) = 
\left(- \frac{G_{\rm F}}{\sqrt{2}} V_{us} V_{ud}\right) \times \sum_{I=1}^6 C_I (\nu) Q_I ( x ; \nu)
.
\ee
The above representation of ${\cal A}^{K \to \pi \ell^+ \ell^-}$
holds at order ${\cal O}(\alpha G_{\rm F})$, $\alpha = e^2/4\pi$ denoting the fine-structure constant, 
and with a corresponding form factor for each $(K, \pi)$ channel, 
that is for the CP conserving transitions $K^+ \to \pi^+ \ell^+ \ell^-$ and $K_S \to \pi^0 \ell^+ \ell^-$, 
but also for the direct-CP violating part of the amplitude for the decay $K_L \to \pi^0 \ell^+ \ell^-$.
The purpose of the perhaps somewhat unfamiliar first expression of ${\cal A}^{K \to \pi \ell^+ \ell^-}$ 
given in Eq. \rf{amplitude} above is to explicitly display the two main components of the weak transition form factor defined
by the second expression in Eq. \rf{amplitude},
\be
W_{K\pi}(z) = W_{K\pi}^{\rm LD}(z ; \nu) + W_{K\pi}^{\rm SD}(z ; \nu)
.
\lbl{SD_LD}
\ee
The first part is dominated by long-distance contributions, the second one is generated 
at short distances, at the electroweak scale or beyond in case new physics sets in at even 
higher energies. 
Notice the appearance of an ultraviolet subtraction scale $\nu$ in each of the two terms
in Eq. \rf{SD_LD}. 
Their sum $W_{K\pi}(z)$ appears in the amplitude ${\cal A}^{K \to \pi \ell^+ \ell^-}$
entering the physical decay rate and should of course not depend on $\nu$ anymore.
Explicitly, one has
\bea
&&
\hspace{-1.5cm}
\left[ s (k + p)_\rho - (M_K^2 - M_\pi^2) (k - p)_\rho \right]
\times \frac{W_{K\pi}^{\rm LD}(z ; \nu)}{16 \pi^2 M_K^2} 
\nonumber\\
&&
= i \! \int \! d^4 x \,
\langle\pi(p)  \vert T \{  j_\rho  (0) {\cal L}^{\Delta S = 1}_{\mbox{\scriptsize non-lept}} (x)  \} \vert K(k) \rangle
\lbl{W_LD}
\eea
for the long-distance dominated part, whereas the contribution from short distances, which
arises from a local term \cite{Witten:1976kx,Gilman:1979ud,Dib:1988md,Dib:1988js}
\be
{\cal L}^{\Delta S = 1}_{\rm lept} (x ; \nu) =
- \frac{G_{\rm F}}{\sqrt{2}} V_{us}^* V_{ud} C_{7V} (\nu)  Q_{7V} (x)
,
\lbl{eff_lag_lept}
\ee
with $Q_{7V} = ( {\bar s}^i d_i )_{V-A} ( {\bar \ell} \ell )_V$, reads
\be
\frac{W_{K\pi}^{\rm SD}(z ; \nu)}{16 \pi^2 M_K^2} = - \left( - \frac{G_{\rm F}}{\sqrt{2}} V_{us} V_{ud} \right)
\times \frac{C_{7V} (\nu)}{4 \pi \alpha} \times C_{K\pi} f_+ (s)
\lbl{W_SD}
\ee
in terms one of the form factor $f_+(s)$ describing the matrix element of the $\Delta S = +1$
neutral current
\bea
&&
\hspace{-1.25cm}
\langle \pi (p) \vert \left( {\bar s} \, \gamma_\rho  d \right) \! (0)  \vert K (k) \rangle
\nonumber\\
&&
=
C_{K\pi} \left[ (k + p )_\rho f_+^{K\pi} (s) + (k-p)_\rho f_-^{K\pi} (s) \right]
.
\lbl{Kl3_ff}
\eea
In these last formulas, $C_{K\pi}$ denotes a Clebsch-Gordan coefficient, chosen such that
$f_+(0)=1$ in the flavour-$SU(3)$ limit.
In both Eqs. \rf{W_LD} and \rf{Kl3_ff} the terms proportional to $(k-p)_\rho$ do not contribute 
to the amplitude ${\cal A}^{K \to \pi \ell^+ \ell^-}$ due to the conservation of the leptonic
current, $(k-p)_\rho \, {\bar{\rm u}} (p_-) \gamma^\rho {\rm v} (p_+) = 0$. For the sake of
completeness, let us recall that in the standard model there is also a contribution to the short-distance 
part of the amplitude ${\cal A}^{K \to \pi \ell^+ \ell^-}$ coming from a term proportional to 
$C_{7A}  Q_{7A} (x)$, with $Q_{7A} = ( {\bar s}^i d_i )_{V-A} ( {\bar \ell} \ell )_A$. 
This term does not play any role in the present discussion, as it does not involve the short-distance
scale $\nu$ and is anyway CKM suppressed. We will therefore not mention it any further.

In $W_{K\pi}^{\rm SD}(z ; \nu)$ the scale dependence is entirely carried by the Wilson coefficient 
$C_{7V} (\nu)$,
\be
\nu \frac{d C_{7V} (\nu)}{d \nu} = \frac{\alpha}{\alpha_s (\nu)} \sum_{J=1}^6 \gamma_{J,7V} (\alpha_s)\,  C_J (\nu) 
.
\lbl{scale_C7V}
\ee
The anomalous dimensions $\gamma_{J,7V} (\alpha_s)$ are known to leading 
\cite{Gilman:1979ud,Dib:1988md,Dib:1988js,Flynn:1988ve} and to next-to-leading 
\cite{Buras:1994qa} orders,
\be
\gamma_{J,7V} (\alpha_s) = \gamma_{J,7V}^{(0)} \, \frac{\alpha_s}{4\pi} 
+ \gamma_{J,7V}^{(1)} \left(\frac{\alpha_s}{4\pi}\right)^2 + \cdots 
.
\ee
In $W_{K\pi}^{\rm LD}(z ; \nu)$, the scale dependence arises from the singular
structure, at short distances, of the time-ordered product of the electromagnetic 
current with the $\Delta S = +1$ effective Lagrangian [both $j^\mu$ and 
${\cal L}_{\mbox{\scriptsize non-lept}}^{\Delta S = +1}$ are finite operators].
This short-distance singularity can be studied perturbatively within the 
operator-product expansion (OPE) \cite{Wilson:1969zs,Wilson:1972ee}. 
After renormalization [we use dimensional regularization in the 
${\overline{\rm MS}}$ subtraction scheme, and below $q^\mu$ denotes a {Euclidean} momentum, 
with $q^2=s<0$, whose components become simultaneously large] one obtains the general structure,
cf. Ref. \cite{DAmbrosio:2018ytt},
\bea
&&
\hspace{-1.35cm}
\lim_{q \to \infty} i
\!\int d^4 x \, e^{i q \cdot x}
T \{ j^\mu (x)
{\cal L}_{\mbox{\scriptsize non-lept}}^{\Delta S = +1} (0) \}  
\nonumber\\
&& \hspace{-1.35cm} 
=
 [ q^\mu q^\rho - q^2 \eta^{\mu\rho} ] \!\times \! {\bar s} \gamma_\rho (1 - \gamma_5) d 
\lbl{OPE}
\\
&& \hspace{-1.35cm} 
\times \! \left(\! - \frac{G_{\rm F}}{\sqrt{2}} V_{us} V_{ud} \right) 
\!\times \! \frac{1}{4 \pi} \sum_{I=1}^6 C_I (\nu) \xi_I (\alpha_s \, ; \nu^2/q^2) 
+ \, {\cal O} (q)
,
\nonumber
\eea
with %[$L(-q^2) \equiv \ln(-\mu^2/q^2)$]
\be
\xi_I (\alpha_s \, ;  \nu^2/s) =
\sum_{p\ge 0} \sum_{r =0}^{p+1}
\xi^I_{pr}
\alpha_s^p (\nu) \ln^r (-\nu^2/s)
.
\lbl{xi_I}
\ee
The subleading terms in the OPE have finite coefficients, and do therefore not
depend on the renormalization scale $\nu$. Consequently [recall that $z\equiv s/M_K^2$],
\bea
\hspace{-0.65cm}
\nu \frac{d W_{K\pi}^{\rm LD}(z ; \nu)}{d\nu}
\!\!&=&\!\! - 16 \pi^2 M_K^2 C_{K\pi} f_+ (s) \times \left( - \frac{G_{\rm F}}{\sqrt{2}} V_{us} V_{ud} \right) 
\nonumber\\
&& \!\!
\times \frac{1}{4 \pi} \nu \frac{d}{d\nu} \sum_{I=1}^6 C_I (\nu) \xi_I (\alpha_s \, ; \nu^2/s)
.~~~
\eea
From Eqs. \rf{OPE} and \rf{xi_I} one infers that in the {Euclidean} region $z\to - \infty$
the form factor $W_{K\pi}^{\rm LD}(z ; \nu)$ behaves asymptotically like
powers of $\ln (-z)$ times powers of the strong coupling $\alpha_s$.

In the present Letter, we wish to discuss how the above short-distance behaviour can be reproduced,
up to the order ${\cal O} (\alpha_s)$, by a model involving an infinite number of equally-spaced 
[in mass squared] zero-width resonances. Models of this kind were considered in various contexts
in the past, see for instance the articles \cite{Shifman:2000jv,Golterman:2001nk,Golterman:2001pj} and references therein. 
Such Regge-type models find their justification in the properties of the QCD spectrum in the limit 
of a large number of {colors} \cite{tHooft74,tHooft:1974pnl,Witten79}. More recently quite efficient methods,
based on the Converse Mapping Theorem \cite{FGD95} and the notion of harmonic sums, were developed in order
to handle such models \cite{Friot:2005cu,Rafael:2012sy,Greynat:2013cta}. 
In the case at hand, the general idea put forward in Ref. \cite{DAmbrosio:2018ytt}
consists in decomposing the long-distance dominated part of the form factor as a sum
\be
W_{K\pi}^{\rm LD} (z) = W_{K\pi}^{\pi\pi} (z) + W_{K\pi}^{\rm res} (z ; \nu) 
.
\lbl{model}
\ee
The first term describes the contribution from the resonant $P$-wave
two-pion intermediate state to $W_{K\pi}(z)$. It is constructed upon assuming 
that it is given by an unsubtracted dispersion integral. The absorptive part consists 
of the two-pion spectral density $\rho_{K\pi}^{\pi\pi} (s)$,
and is obtained upon inserting a two-pion intermediate state in the 
representation of the form factor given in Eq. \rf{amplitude}. This contribution
is not relevant for the discussion of the short-distance properties, and we
refer the interested reader to Ref. \cite{DAmbrosio:2018ytt} for details. 
The second term describes the contributions from the intermediate 
states with higher thresholds as a sum of zero-width resonances. 
The weight with which each resonance contributes must be chosen such as to reproduce
the short-distance behaviour given in Eqs. \rf{OPE} and \rf{xi_I}. In Ref. \cite{DAmbrosio:2018ytt},
this matching has been achieved at order ${\cal O} (\alpha_s^0)$, which involves only
a constant term and a term linear in $\ln (-z)$. Here we wish to extend this matching to
the order ${\cal O} (\alpha_s)$, where we also encounter a term quadratic in $\ln (-z)$.

The remainder of this Letter is thus organized as follows. In section \ref{SD_coeffs}
we determine the required coefficients $\xi_{01}^I$, $\xi_{11}^I$ and $\xi_{12}^I$, relying
on a renormalization-group argument given in Ref. \cite{DAmbrosio:2018ytt}. For
{pedagogical} reasons, we then review (Section \ref{matching_LO}) the matching at order ${\cal O} (\alpha_s^0)$,
making the discussion of Ref. \cite{DAmbrosio:2018ytt} simpler and, hopefully, also
more intuitive. The matching at order ${\cal O} (\alpha_s)$ is then presented in Section \ref{matching_NLO}.
Finally, we discuss some consequences and features of our results (Section \ref{discussion}).

\section{Determination of the short-distance coefficients}\label{SD_coeffs}
\setcounter{equation}{0}

\begin{figure*}[t]
\center\epsfig{figure=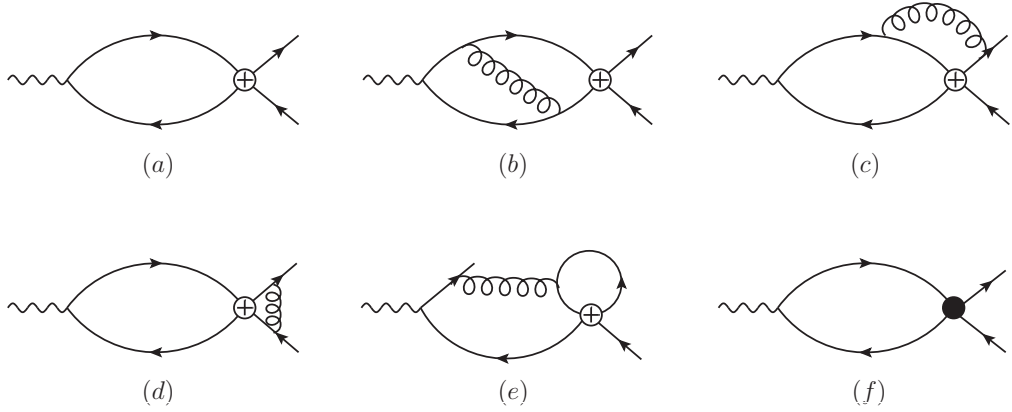,height=5.5cm}
\caption{Diagrams contributing to the leading
short-distance behaviour of the time-ordered product of the 
electromagnetic current (materialized by the wiggly line on the left)
with one of the four-quark operators, whose insertion is
shown by the crossed-circle ($\oplus$) vertex on the right of diagram $(a)$,
of order ${\cal O}(\alpha_s^0)$, and of the diagrams
$(b)$ to $(e)$. The latter also form a representative subset of
the diagrams contributing to the gluonic corrections at order ${\cal O}(\alpha_s)$.
The vertex represented by a circular black blob ({\large$\bullet$}) on the right of diagram $(f)$ corresponds
to the insertion of an order ${\cal O}(\alpha_s)$
{counterterm} into the lowest-order diagram $(a)$. The other external lines
represent an incoming $d$ quark and an outgoing $s$ quark, while the
loop consists of a  ${\bar q}$-$q$ pair, where $q=u,d,s$.} \label{OPE_QCD_fig}
\end{figure*}

From the perspective of perturbative QCD, the matrix element that defines the long-distance
dominated part $W_{K\pi}^{\rm LD}(z ; \nu)$ is described by the {Feynman} diagrams of the
type shown {\rm in} Fig. \ref{OPE_QCD_fig}. This is obviously not a realistic description
of $W_{K\pi}^{\rm LD}(z ; \nu)$, except at short distances, when the momentum transfered
to the electromagnetic current becomes large in the space-like or Euclidean domain, $s\to -\infty$.
The leading contribution to the OPE of the electromagnetic current and the four-quark operators 
describing $\Delta S = +1$ transitions involves the neutral-current operator ${\bar s} \gamma_\rho (1 - \gamma_5) d$,
and corresponds to the diagram $(a)$ of Fig. \ref{OPE_QCD_fig}. The calculation with a bare
four-quark operator $Q_I$ is straightforward, see Ref. \cite{DAmbrosio:2018ytt}, and gives
\bea
&&
\hspace{-1.35cm}
\lim_{q \to \infty} i
\!\int d^4 x \, e^{i q \cdot x}
T \{ j^\mu (x)
{\cal L}_{\mbox{\scriptsize non-lept}}^{\Delta S = +1} (0) \}  
\nonumber\\
&& \hspace{-1.35cm} 
= [ q^\mu q^\rho - q^2 \eta^{\mu\rho} ] \times [{\bar s} \gamma_\rho (1 - \gamma_5) d](0)
\times \left( - \frac{G_{\rm F}}{\sqrt{2}} V_{us} V_{ud} \right) 
\nonumber\\
&& \hspace{-1.35cm}
\times \frac{1}{4 \pi} \sum_{I=1}^6 C_I \left\{ \xi_{00}^I 
- \xi_{01}^I \left[ \frac{2}{D-4} + \ln\left(\frac{- q^2}{\nu^2}\right) \right]
+ {\cal O}(\alpha_s) \right\} 
\nonumber\\
&& \hspace{-1.35cm}
+ \ {\cal O} (q)
.
\eea
with
\bea
\xi_{01}^I &=& \frac{1}{4\pi} \, \frac{8}{9} 
\times
\left(
N_c \, , \, 1 \, , \, -1 \, , \, -N_c \, , \, 0 \, , \, 0
\right),
\nonumber\\
\xi_{00}^I &=& \frac{\xi_{01}^I}{3} \times 
\left\{
\begin{array}{l}
 2~~ {\rm NDR} \\
 \\
 5~~ {\rm HV}
\end{array}
\right. 
,
\lbl{xi_LO}
\eea

\noindent
The result for $\xi_{00}^I$ depends on the scheme used to handle the Dirac matrices 
in $D$ dimensions, here either naive dimensional regularization (NDR) 
\cite{Chanowitz:1979zu} or the 't Hooft-Veltman (HV) scheme \cite{tHooft:1972tcz,Breitenlohner:1977hr}.
The divergent part is removed through the renormalization of the (bare) coupling $C_{7V}$.
Notice also that at this order the Wilson coefficients are not yet running.
In the presence of QCD corrections, the four-quark
operators and the corresponding Wilson coefficients are renormalized. But this
does not take care of all the divergences, since two external lines of the $Q_I$ operators
are closed into a loop with the insertion of the electromagnetic current, see Fig. \ref{OPE_QCD_fig}. 
These remaining divergences are again absorbed through the higher-order renormalization
of $C_{7V}$. To all orders in the powers of $\alpha_s$, and after renormalization 
through minimal subtraction in the ${\overline{\rm MS}}$ scheme, the leading term in the OPE 
then takes the form given in Eqs. \rf{OPE} and \rf{xi_I}.
From the scale dependence of the Wilson coefficients $C_{7V}(\nu)$, given in Eq. \rf{scale_C7V},
and $C_I(\nu)$,
\be
\nu \frac{d C_I (\nu)}{d \nu} = \sum_{J=1}^6 \gamma_{J,\, I} (\alpha_s) C_J (\nu)
,
\lbl{scale_CI}
\ee
one infers that the total form factor in Eq. \rf{SD_LD} will be scale independent provided the equation
\be
\nu \frac{d \xi_I (\alpha_s \, ; \nu^2/s)}{d\nu} + \sum_{J=1}^6 \gamma_{I,J} (\alpha_s) \xi_J (\alpha_s \, ; \nu^2/s)
= - \frac{\gamma_{I,7V} (\alpha_s)}{\alpha_s (\nu)} 
\lbl{RG_equ}
\ee
%\bea
%&&
%\hspace{-1.25cm}
%\nu \frac{d \xi_I (\alpha_s \, ; \nu^2/s)}{d\nu} + \sum_{J=1}^6 \gamma_{I,J} (\alpha_s) \xi_J (\alpha_s \, ; \nu^2/s)
%= - \frac{\gamma_{I,7V} (\alpha_s)}{\alpha_s (\nu)} 
%\nonumber\\
%&&
%= - \frac{\gamma_{I,7V} (\alpha_s)}{\alpha_s (\nu)} 
%\lbl{RG_equ}
%\eea
holds. At order ${\cal O}(\alpha_s^0)$, this allows to recover
the values of the coefficients $\xi_{01}^I$ from the known one-loop anomalous 
dimension matrices. 
Including first-order QCD corrections, one obtains, after renormalization,
\bea
&&
\hspace{-1.25cm}
\lim_{q \to \infty} i
\!\int d^4 x \, e^{i q \cdot x}
T \{ j^\mu (x)
{\cal L}_{\mbox{\scriptsize non-lept}}^{\Delta S = +1} (0) \} 
\lbl{OPE_NLO}
\\
&& \hspace{-1.0cm}
= [ q^\mu q^\rho - q^2 \eta^{\mu\rho} ] \times {\bar s} \gamma_\rho (1 - \gamma_5) d
\times \left( - \frac{G_{\rm F}}{\sqrt{2}} V_{us} V_{ud} \right) 
\nonumber\\
&& \hspace{-1.0cm}
\times \frac{1}{4 \pi} \sum_{I=1}^6 C_I (\nu) \left\{ \xi_{00}^I - \xi_{01}^I \ln\left(\frac{- q^2}{\nu^2}\right)
+ \alpha_s (\nu) \left[ \xi_{10}^I
\right.\right.\nonumber\\
&& \hspace{-1.0cm}
\left.\left.
- \xi_{11}^I \ln\left(\frac{- q^2}{\nu^2}\right) 
+ \xi_{12}^I \ln^2\left(\frac{- q^2}{\nu^2}\right) \right] + {\cal O}(\alpha_s^2) \right\}
+ {\cal O} (q)
.
\nonumber
\eea
From the above renormalization-group argument and the known two-loop 
anomalous dimension matrices \cite{Altarelli:1980fi,Buras:1989xd,Buras:1991jm,Buras:1992tc,Ciuchini:1993vr}, 
one infers the relations
\be
\xi_{12}^I = \frac{1}{(4\pi)^2} \, \frac{4}{27}  \left(N_c - \frac{1}{N_c} \right)
\times
\left(
0 \, , \, - 8 \, , \, + 11 \, , \, N_f \, , \, 0 \, , \, N_f
\right) \!,
\ee
and
\bea
&& \hspace{-1.25cm}
\xi^I_{11} = \frac{1}{(4\pi)^2} \, \frac{8}{3}  \left(N_c - \frac{1}{N_c} \right)
\nonumber\\
&& \hspace{-0.75cm}
\times
\left\{
\begin{array}{l}
\!\!\left( \frac{N_c}{2} \, , \, -\frac{19}{18} \, , \, \frac{17}{9} \, , \, \frac{7}{6} - \frac{N_c}{2} \, , \, 0 \, , \, \frac{7}{6} \right)~~ {\rm NDR} \\
\\
\!\!\left( \frac{N_c}{2} \, , \, -\frac{5}{18} \, , \, \frac{13}{9} \, , \, \frac{7}{6} - \frac{N_c}{2} \, , \, 0 \, , \, \frac{7}{6} \right)~~ {\rm HV}
\end{array}
\right. 
\!.
\eea
The coefficients $\xi_{10}^I$ cannot be obtained this way, and their
determination would require a full two-loop calculation, which we will 
however not need to attempt here.

\section{The matching at order ${\cal O} (\alpha_s^0)$}\label{matching_LO}
\setcounter{equation}{0}

 It is convenient to represent the resonance contribution to the form factor in
 Eq. \rf{model} as a dispersive integral
\be
W^{\rm res}_{K\pi} (z ; D) = \frac{C_{K\pi} f_+^{K\pi} (z M_K^2)}{4\pi} \int dx \, \frac{\rho^{\rm res}_{K\pi}(x ; D)}{x - z M_K^2 - i0}
,
\ee
where the spectral density is constructed order by order in the expansion in powers
of the strong coupling $\alpha_s$,
\be
\rho^{\rm res}_{K\pi}(x ; D) = \rho^{\rm res;0}_{K\pi}(x ; D) + \rho^{\rm res;1}_{K\pi}(x ; D) 
+ \cdots ,
\ee
and is defined in $D$ dimensions. This last point requires some explanation.
Indeed, by power counting the naive superficial degree of divergence of the diagrams in Fig. \ref{OPE_QCD_fig}
is two, and becomes actually zero once the Ward identity following from the conservation of the electromagnetic current
is implemented. Therefore, in four dimensions the form factor satisfies a once-subtracted dispersion relation, which
would thus constitute an appropriate representation to start with. However, the information from 
short distances at our disposal, and summarized in the preceding section, comes from calculations 
done within a dimensional renormalization scheme with minimal subtraction and not 
within a momentum subtraction one. Since we want to make direct use of this information without transforming it 
first into a different scheme, we choose instead to start from an unsubtracted dispersion relation in $D$ dimensions. 
As will hopefully become clear in the sequel, far from being an unnecessary complication, this choice even presents some advantages
in actually guiding our intuition in the process of constructing an appropriate ansatz for the spectral density
$\rho^{\rm res;0}_{K\pi}(x ; D)$ or $\rho^{\rm res;1}_{K\pi}(x ; D)$.

At the one-loop level, i.e. order ${\cal O} (\alpha_s^0)$, the resonance representation we are looking for
should match, at large negative values of $s$, the behaviour in the same limit
of the unrenormalized QCD diagram $(a)$ of Fig. \ref{OPE_QCD_fig}. In order to reproduce a logarithmic 
behaviour at short distances, an infinite set of resonances is required \cite{Witten79}. We will consider a simple  
Regge-type description of the resonance spectrum in terms of equally spaced [in mass squared] zero-width states.
Accordingly, the general structure of $\rho^{\rm res;0}_{K\pi}(x ; D)$ is given by
\bea
\hspace{-0.5cm}
\rho^{\rm res;0}_{K\pi}(x ; D) \!\!&=&\!\! A_0 (D) (4\pi)^{2 - \frac{D}{2}} \left( \frac{M^2}{\nu^2_{\rm MS}} \right)^{\frac{D}{2} - 2}
\Gamma \left(2 - \frac{D}{2} \right) 
\nonumber\\
&& \!\! \times\sum_{n\ge 1} M^2 \mu_n^{(0)} (D) \delta (x - nM^2)
,
\eea
with so-far unspecified functions $A_0(D)$ and $\mu_n^{(0)} (D)$. The additional prefactors simply 
account for the structure of diagram $(a)$ of Fig. \ref{OPE_QCD_fig} in $D$ dimensions. Since the one-loop 
divergence is already contained in the factor $\Gamma (2 - D/2)$, $A_0(D)$ has to be regular at $D=4$. 
It accounts for the scheme dependence, for instance in handling Dirac matrices
in $D$ dimensions, see Section \ref{SD_coeffs}. $M$ denotes
the mass of the lowest-lying resonance and $\nu_{\rm MS}$ denotes the renormalization scale in the 
minimal subtraction scheme.
Then, with $w\equiv -s/M^2$,
\bea
\hspace{-0.5cm}
\int dx \, \frac{\rho^{\rm res;0}_{K\pi}(x ; D)}{x+wM^2} \!\!&=& \!\!
A_0 (D) (4\pi)^{2 - \frac{D}{2}} \left( \frac{M^2}{\nu^2_{\rm MS}} \right)^{\frac{d}{2} - 2}
\nonumber\\
&&\!\! \times \,
\Gamma \left(2 - \frac{D}{2} \right) 
\sum_{n\ge 1} \frac{\mu_n^{(0)} (D)}{n+w}
.
\eea
The value of the above dispersive integral at $w=0$ reproduces the divergent part of
the diagram $(a)$ of Fig. \ref{OPE_QCD_fig} 
%[$\Gamma(2-D/2) =  -2/(D-4) - \gamma_E + {\cal O} (D-4)$, where $\gamma_E$ denotes the Euler constant]
\bea
\hspace{-0.5cm}
\int \frac{dx}{x} \, \rho^{\rm res;0}_{K\pi}(x ; D) \!\!\!&=&\!\!\! 
A_0 (D) (4\pi)^{2 - \frac{D}{2}} \left( \frac{M^2}{\nu^2_{\rm MS}} \right)^{\frac{d}{2} - 2} 
\nonumber\\
&&
\!\!\! \times \Gamma \left(2 - \frac{D}{2} \right)
\sum_{n\ge 1} \frac{\mu_n^{(0)} (D)}{n}
.
\eea
$A_0(4)$ being so far a free parameter, one may, without loss of generality, require that
\be
\sum_{n\ge 1} \frac{\mu_n^{(0)} (D)}{n} = 1 + (D-4) \mu^{(0)} + {\cal O} ((D-4)^2)
,
\lbl{cond_1}
\ee
with $\mu^{(0)}$ a constant that remains unspecified for the time being.
Then
\bea
\hspace{-0.55cm}
\int \frac{dx}{x} \, \rho^{\rm res;0}_{K\pi}(x ; D) \!\!&=& \!\!
A_0(4) \bigg[
- \frac{2}{D-4} + \ln \frac{\nu^2}{M^2} 
\\
&& \!\!
- 2 \frac{A_0^\prime(4)}{A_0(4)}
- 2 \mu^{(0)}
\bigg] + {\cal O} (D-4),
\nonumber
\eea
with $\nu \equiv \nu_{\rm MS} e^{-\gamma_E/2} \sqrt{4\pi}$
the subtraction scale in the ${\overline{\rm MS}}$ scheme. 
The constant $\mu^{(0)}$ can actually be absorbed without loss
of generality into $A_0^\prime(4)$, which has also not been specified so far.
The remaining, subtracted, dispersive integral
\bea
\hspace{-0.65cm}
w M^2 \! \int \frac{dx}{x} \, \frac{\rho^{\rm res;0}_{K\pi}(x ; D)}{x+wM^2} \!\!&=& \!\! 
A_0 (D) (4\pi)^{2 - \frac{D}{2}} \left( \frac{M^2}{\nu^2_{\rm MS}} \right)^{\frac{d}{2} - 2} 
\\
&& \times \,
\Gamma \left(2 - \frac{D}{2} \right) \sum_{n\ge 1} \frac{w \mu_n^{(0)} (D)}{n(n+w)}
\nonumber
\eea
should then be finite as $D\to 4$. This in turn will be the case if we require
\be
\mu_n^{(0)} (D) = (D-4) {\bar\mu}^{(0)}_n + {\cal O} ((D-4)^2) 
\lbl{cond_2}
\ee
and 
\be
\xi(w) = \sum_{n\ge 1} \frac{{\bar\mu}_n^{(0)}}{n(n+w)} ~~ {\rm converges}
.
\lbl{cond_3}
\ee
Consequently
\be
w M^2 \int \frac{dx}{x} \, \frac{\rho^{\rm res;0}_{K\pi}(x ; D)}{x+wM^2} = 
A_0 (4) (- 2 w) \xi (w) + {\cal O} (D-4)
.
\ee
Finally, in order to reproduce the correct matching with the short-distance behaviour,
one must also require 
\be
- 2 w \xi (w) = \ln w + {\rm Cst} + \cdots
\ee 
as $w \to +\infty$. At this stage, one may observe that ${\mu}_n^{(0)} \propto (D-4) {\bar\mu}^{(0)}$, 
with ${\bar\mu}^{(0)}$ a constant,
would provide a convergent series $\xi (w)$, endowed with an asymptotic logarithmic behaviour. 
But it would fail to satisfy the condition \rf{cond_1}. As we now show, this defect can be easily
repaired. Intuitively, the task will consist in providing a convergence factor for the sum \rf{cond_1} when $D<4$, 
but which is no longer operative for $D=4$, otherwise the sum $\xi(w)$ would converge too quickly,
and would no longer exhibit a logarithmic behaviour for large values of $w$. Instead, it would rather behave 
as a constant or an inverse power of $w$. 
Consider therefore the ansatz
\be
\mu_n^{(0)} (D) = {\rm a}^{(0)} (D) n^{\frac{D}{2} - 2}
.
\ee
Then one has
\be
\sum_{n\ge 1} \frac{\mu_n^{(0)} (D)}{n} = {\rm a}^{(0)} (D) \zeta \left( 3 - \frac{D}{2}  \right)
,
\ee
where $\zeta(s)$ denotes Riemann's zeta-function, see \cite[section 25.2]{DLMF}.
Since, as $D\to 4$, $\zeta(3-D/2) \sim -2/(D-4) + \gamma_E$, where $\gamma_E$ is 
the Euler constant, any choice of the form
\be
{\rm a}^{(0)} (D) = \frac{1}{\zeta \left( 3 - \frac{D}{2}  \right) + f(D)}
,
\ee
where $f(D)$ is an arbitrary function regular at $D=4$, will lead to
\be
\sum_{n\ge 1} \frac{\mu_n^{(0)} (D)}{n} = 1 + \frac{f(4)}{2} (D-4) + {\cal O}((D-4)^2)
\ee
and thus satisfy the condition \rf{cond_1} with $\mu^{(0)} = f(4)/2$. 
The condition \rf{cond_2} is then also satisfied, with ${\bar\mu}_n = - 1/2$, so that
\be
\Gamma \left(2 - \frac{D}{2} \right) \mu_n (D) = 1 + {\cal O} (D-4)
.
\ee
The condition \rf{cond_3} is met as well, with
\be
- 2 w \xi (w) = \sum_{n\ge 1} \frac{w}{n(n+w)} = \gamma_E + \psi(1+w)
.
\ee
Putting everything together, one ends up with 
\bea
&&
\hspace{-1.25cm}
\int dx \, \frac{\rho^{\rm res;0}_{K\pi}(x ; D)}{x+wM^2} =
A_0(4) \bigg\{
- \frac{2}{D-4} + \ln \frac{\nu^2}{M^2} - 2 \frac{A_0^\prime(4)}{A_0(4)}
\nonumber\\
&&
-2 \mu^{(0)} - \left[ \gamma_E + \psi (1 + w) \right]
\bigg\}  + {\cal O}(D-4)
.
\eea
In this last formula, the di-gamma function $\psi$ resums the dispersive integral
\be
\psi(1+w) = - \gamma_E \!+\! M^2 w \int \! \frac{dx}{x} \, \frac{1}{x + M^2 w} \sum_{n\ge 1} M^2 \delta (x - nM^2)
.
\ee
Finally, for large positive $w$ the di-gamma function behaves as $\psi(1+w) \sim \ln w = \ln(-s/M^2)$, so that the correct 
short-distance behaviour is also recovered, provided one takes
\be
A_0 (4) = - 16 \pi^2 M_K^2 \left( - \frac{G_{\rm F}}{\sqrt{2}} V_{us} V_{ud} \right)
\times \sum_{I=1}^6 C_I (\nu) \xi_{01}^I
\ee
and
\bea
\hspace{-0.67cm}
A_0^\prime(4) + \left( \mu^{(0)} + \frac{\gamma_E}{2} \right) A_0(4) \!\!\!&=&\!\!\!
+ 16 \pi^2 M_K^2 \! \left( - \frac{G_{\rm F}}{\sqrt{2}} V_{us} V_{ud} \right)
\nonumber\\
&& \!\!\!
\times \sum_{I=1}^6 C_I (\nu) \frac{\xi_{00}^I}{2}
.
\eea
Minimal subtraction of the divergence leads to the renormalized dispersion relation
\bea
&&
\hspace{-1.37cm}
\int dx \, \frac{\rho^{\rm res;0}_{K\pi}(x ; D)}{x+wM^2} =
- 16 \pi^2 M_K^2 \! \left( \! - \frac{G_{\rm F}}{\sqrt{2}} V_{us} V_{ud} \right)
\!\times \! \sum_{I=1}^6 C_I (\nu)
\nonumber\\
&&
\hspace{+1.5cm}
\times\left[
\xi_{01}^I \left(\ln \frac{\nu^2}{M^2} - \psi(1+w) \right) + \xi_{00}^I
\right]
\nonumber\\
&&
\hspace{+1.5cm}
+ \, {\cal O} (D-4)
,
\lbl{ren_disp_rho_0}
\eea
with
\bea
\hspace{-0.65cm}
\rho^{\rm res;0}_{K\pi}(x ; D) \!\!&=& \!\! - 16 \pi^2 M_K^2 \left( - \frac{G_{\rm F}}{\sqrt{2}} V_{us} V_{ud} \right)
\times \sum_{I=1}^6 C_I (\nu) 
\nonumber\\
&& \!\!
\times
\bigg\{ 
\xi_{01}^I - \frac{D-4}{2} \left[\xi_{00}^I + (f(4) + 2 \gamma_E) \xi_{01}^I \right] \! \bigg\}
\nonumber\\
&& \!\!
\times
\left( \frac{M^2}{\nu^2} \right)^{\frac{D}{2} - 2}
\Gamma \left(2 - \frac{D}{2} \right) 
\lbl{rho_0}
\\
&& \!\!
\times
\sum_{n\ge 1} \frac{M^2 n^{\frac{D}{2} - 2}}{\zeta \left( 3 - \frac{D}{2} \right) + f(D)} \, \delta (x - nM^2)
.
\nonumber
\eea

We may now, in some sense, reverse-engineer the whole construction, starting directly from the representation 
of $\rho^{\rm res;0}_{K\pi}(x ; D)$ in $D$ dimensions given in Eq. \rf{rho_0} above and showing that it satisfies 
the required properties. This will be useful in Section \ref{matching_NLO}, where we will only sketch the 
construction of the spectral density $\rho^{\rm res;0}_{K\pi}(x ; D)$, give the result, and show that it indeed
exhibits the appropriate features.
Using the Mellin representation
\be
\frac{1}{1+\frac{w}{n}} = \int_{c_1-i\infty}^{c_1+i\infty} \frac{du}{2 \pi i} \left(\frac{w}{n}\right)^{-u} \Gamma(u) \Gamma(1-u)
,
\lbl{mellin}
\ee
where $0< c_1 <1$, one has
\bea
&&
\hspace{-1.25cm}
\int dx \, \frac{\rho^{\rm res;0}_{K\pi}(x ; D)}{x+wM^2}
\nonumber\\
&&
\hspace{-1.25cm}
= - 16 \pi^2 M_K^2 \left( - \frac{G_{\rm F}}{\sqrt{2}} V_{us} V_{ud} \right)
\times \sum_{I=1}^6 C_I (\nu)  \bigg\{ 
\xi_{01}^I - \frac{D-4}{2}
\nonumber\\
&&
\hspace{-0.9cm} 
\times \big[\xi_{00}^I + (f(4) + 2 \gamma_E) \xi_{01}^I \big] \bigg\}
\times
\left( \frac{M^2}{\nu^2} \right)^{\frac{D}{2} - 2}
\nonumber\\
&&
\hspace{-0.9cm} 
\times
\frac{\Gamma \left(2 - \frac{D}{2} \right)}{\zeta \left( 3 - \frac{D}{2} \right) + f(D)} \times
\int_{c_1-i\infty}^{c_1+i\infty} \frac{du}{2 \pi i} w^{-u} \Gamma(u) \Gamma(1-u) 
\nonumber\\
&&
\hspace{-0.9cm} 
\times
\zeta \left( 3 - \frac{D}{2} -u \right)
.
\eea
The first singularity of the integrand lying on the left of the fundamental band $c_1={\rm Re} \, u \in ]0,2-D/2[$ 
is a simple pole at $u=0$, coming from the factor $\Gamma(u)$. According to the Converse Mapping Theorem
\cite{FGD95} one therefore has
\bea
&&
\hspace{-1.25cm}
\int dx \, \frac{\rho^{\rm res;0}_{K\pi}(x ; D)}{x+wM^2}
\nonumber\\
&&
\hspace{-1.25cm}
\underset{w \rightarrow 0}{=}
- 16 \pi^2 M_K^2 \!\left( \! - \frac{G_{\rm F}}{\sqrt{2}} V_{us} V_{ud} \right)
\times \sum_{I=1}^6 C_I (\nu)  \bigg\{ 
\xi_{01}^I - \frac{D-4}{2}
\nonumber\\
&&
\hspace{-0.6cm} 
\times \big[\xi_{00}^I + (f(4) + 2 \gamma_E) \xi_{01}^I \big] \bigg\}
\times
\left( \frac{M^2}{\nu^2} \right)^{\frac{D}{2} - 2}
\nonumber\\
&&
\hspace{-0.6cm} 
\times
\frac{\Gamma \left(2 - \frac{D}{2} \right) \zeta \left( 3 - \frac{D}{2} \right)}{\zeta \left( 3 - \frac{D}{2} \right) + f(D)} 
\nonumber\\
&&
\hspace{-1.25cm} 
\underset{D \rightarrow 4}{=} 
- 16 \pi^2 M_K^2 \!\! \left( - \frac{G_{\rm F}}{\sqrt{2}} V_{us} V_{ud} \right)
\times \sum_{I=1}^6 C_I (\nu)  \bigg\{ 
\frac{-2}{D-4} \xi_{01}^I
\nonumber\\
&&
\hspace{-0.6cm} 
+ \, \xi_{00}^I + \xi_{01}^I \gamma_E + \xi_{01}^I \ln\frac{M^2}{\nu^2} \bigg\}
+ {\cal O} (D-4)
.
\eea
The first singularity of the integrand lying on the right of the fundamental band $c_1={\rm Re} \, u \in ]0,2-D/2[$ 
is a {simple} pole occurring at $u=2-D/2$, with $\zeta \left( 3 - \frac{D}{2} -u \right) \sim -1/(u-2+D/2)$, 
so that, according to the Converse Mapping Theorem,
\bea
&&
\hspace{-1.25cm} 
\int dx \, \frac{\rho^{\rm res;0}_{K\pi}(x ; D)}{x+wM^2}
\nonumber\\
&&
\hspace{-1.25cm} 
\underset{w \rightarrow +\infty}{\sim} 
- 16 \pi^2 M_K^2 \left( - \frac{G_{\rm F}}{\sqrt{2}} V_{us} V_{ud} \right)
\times \sum_{I=1}^6 C_I (\nu)  \bigg\{ 
\xi_{01}^I
\nonumber\\
&&
\hspace{-0.35cm}  
- \, \frac{D-4}{2} \left[\xi_{00}^I + (f(4) + 2 \gamma_E) \xi_{01}^I \right] \bigg\} 
\times
\left( \frac{M^2}{\nu^2} \right)^{\frac{D}{2} - 2}
\nonumber\\
&&
\hspace{-0.35cm}
\times \,
\frac{\Gamma \left(2 - \frac{D}{2} \right)}{\zeta \left( 3 - \frac{D}{2} \right) + f(D)} 
\times \Gamma \left(2 - \frac{D}{2} \right) \Gamma \left(\frac{D}{2} - 1 \right) 
\nonumber\\
&&
\hspace{-0.35cm}
\times \,
\left[ w^{\frac{D}{2} - 2} + {\cal O} \left( w^{-1} \right) \right]
.\qquad~
\eea
Considering next the limit $D\to 4$, one indeed recovers the expected result:
\bea
&&
\hspace{-1.25cm} 
\int dx \, \frac{\rho^{\rm res;0}_{K\pi}(x ; D)}{x+wM^2} 
\nonumber\\
&&
\hspace{-1.25cm} 
\underset{w \rightarrow +\infty}{\sim} 
- 16 \pi^2 M_K^2 \left( - \frac{G_{\rm F}}{\sqrt{2}} V_{us} V_{ud} \right)
\times \sum_{I=1}^6 C_I (\nu) 
\\
&&
\hspace{-0.35cm} 
\times \,  \bigg\{ 
\frac{-2}{D-4} \xi_{01}^I + \xi_{00}^I + \xi_{01}^I \ln\frac{\nu^2}{-s} \bigg\}
+ {\cal O}(D-4)
.
\nonumber
\eea
Summarizing the preceding analysis, we find {indeed} that the spectral density \rf{rho_0} 
reproduces the minimally subtracted dispersive integral in Eq. \rf{ren_disp_rho_0}.
Let us notice, at this stage, that the result \rf{ren_disp_rho_0} for $D=4$ does no 
longer depend at all on $f(D)$. We will return to this point in Section \ref{discussion} below.

\section{The matching at order ${\cal O} (\alpha_s)$}\label{matching_NLO}
\setcounter{equation}{0}

In order to include the ${\cal O}(\alpha_s)$ corrections that arise at two loops, it is necessary to start with
a somewhat more involved expression of the spectral density. We again let ourselves be guided by the perturbative 
structure of this contribution. On the basis of the order ${\cal O}(\alpha_s)$ diagrams shown in Fig. \ref{OPE_QCD_fig}, 
we are led to consider as a starting point the sum of two terms,
\be
\rho^{\rm res;1}_{K\pi}(x ; D) = \rho^{\rm res;1a}_{K\pi}(x ; D) + \rho^{\rm res;1b}_{K\pi}(x ; D)
\ee
with
\bea
&& \hspace{-1.3cm}
\rho^{\rm res;1a}_{K\pi}(x ; D) = A_{1a} (D) (4\pi)^{4 - D} \!\left( \frac{M^2}{\nu^2_{\rm MS}} \right)^{D-4}
\! \left[\Gamma \left(2 - \frac{D}{2} \right)\right]^2
\nonumber\\
&& \hspace{0.7cm}
\times \sum_{n\ge 1} M^2 \mu_n^{(1a)} (D) \, \delta (x - nM^2)
,
\nonumber\\
&& \hspace{-1.3cm}
\rho^{\rm res;1b}_{K\pi}(x ; D) =
\frac{A_{1b} (D)}{D-4} \, (4\pi)^{2 - \frac{D}{2}} \left( \frac{M^2}{\nu^2_{\rm MS}} \right)^{\frac{D}{2} - 2}
\Gamma \left(2 - \frac{D}{2} \right)
\nonumber\\
&& \hspace{0.7cm}
\times \sum_{n\ge 1} M^2 \mu_n^{(1b)} (D) \, \delta (x - nM^2)
,
\eea
where both functions $A_{1a} (D)$ and $A_{1b} (D)$ are regular at $D=4$.
The first term, $\rho^{\rm res;1a}_{K\pi}(x ; D)$, corresponds to the genuine two-loop diagrams,
like the graphs $(b)$ to $(e)$ of Fig. \ref{OPE_QCD_fig}. The second term, $\rho^{\rm res;1b}_{K\pi}(x ; D)$, 
stands for the diagram $(f)$, i.e. the one-loop graph $(a)$ with the insertion of a one-loop counterterm 
proportional to $1/(D-4)$. Without loss of generality, one may impose a normalization
condition like \rf{cond_1} for each set of coefficients separately, and from there proceed as in 
Section \ref{matching_LO}. We will not go through the details of this straightforward exercise, but rather
state the final result and show that it satisfies the required properties. Before that, let us make a few
useful remarks. 

With the normalizations set as in Eq. \rf{cond_1} for both $\rho^{\rm res;1a}_{K\pi}(x ; D)$
and $\rho^{\rm res;1b}_{K\pi}(x ; D)$, the corresponding dispersive integrals at $w=0$ display double and 
{ simple} poles at $D=4$. The { simple} poles contain a contribution proportional to $\ln (M^2/\nu^2)$, which, 
on general grounds \cite{Collins74}, is not allowed, and hence has to cancel in the sum of the two dispersive integrals.
For this to happen, we need to impose the condition $A_{1b} (4) = 4 A_{1a} (4)$.

Next, $\rho^{\rm res;1b}_{K\pi}(x ; D)$ has to describe the same structure {as} $\rho^{\rm res;0}_{K\pi}(x ; D)$
up to an additional factor $\propto 1/(D-4)$ coming from the inserted counterterm. It is thus natural to consider 
for it the ansatz
\be
\mu_n^{(1b)} (D) = \frac{n^{\frac{D}{2} - 2}}{\zeta\left( 3 - \frac{D}{2} \right) + {\tilde f} (D)}
,
\ee
where ${\tilde f}(D)$ is an arbitrary function regular at $D=4$.
Then the resulting subtracted dispersive integral
\be
w M^2 \int \frac{dx}{x} \, \frac{\rho^{\rm res;1b}_{K\pi}(x ; D)}{x+wM^2} 
\ee
still contains a term proportional to $\psi (1+w)/(D-4)$, which has eventually
to be canceled by a similar contribution from the dispersive integral involving 
$\rho^{\rm res;1a}_{K\pi}(x ; D)$.

Turning to the latter, we consider the ansatz
\be
\mu_n^{(1a)} (D) = \frac{n^{y(D-4)}}{\zeta\left( 1+4y-yD \right) + g (D)}
,
\ee
where $g(D)$ is an arbitrary function regular at $D=4$, and $y$ is a so-far free parameter.
Considering the dispersive integral
\be
w M^2 \int \frac{dx}{x} \, \frac{\rho^{\rm res;1a}_{K\pi}(x ; D)}{x+wM^2} 
,
\ee
we find that the required cancellation of the unwanted pole terms takes place for the choice $y=1$.

Adjusting the remaining free parameters such as to reproduce the asymptotic 
behaviour of Eq. \rf{OPE_NLO}, we finally arrive at the result\footnote{Up to a factor $(-1)^n$, 
the Stieltjes constants $\gamma_n$ give the coefficients
of the Taylor expansion of the regular part of $\zeta (s)$ at $s=1$, with $\gamma_0 = \gamma_E = 0.577216\cdots$,
$\gamma_1=-0.0728\cdots$, $\gamma_2=-0.00969\cdots$; see \cite[section 25.2]{DLMF}.}
\bea
&&
\hspace{-1.25cm}
\rho^{\rm res;1}_{K\pi}(x ; D) 
\nonumber\\
&& 
\hspace{-1.25cm}
= - 16 \pi^2 M_K^2 \left( - \frac{G_{\rm F}}{\sqrt{2}} V_{us} V_{ud} \right) \alpha_s (\nu)
\sum_{I=1}^6 C_I (\nu) 
\nonumber\\
&& 
\hspace{-1.25cm}
\times
\left\{
\left( \frac{M^2}{\nu^2} \right)^{D-4}
\left[ \Gamma \left(2 - \frac{D}{2} \right) \right]^2 e^{\gamma_E(4-D)} \xi_{12}^I
\right.
\nonumber\\ 
&& 
\times \sum_{n\ge 1} M^2
\frac{n^{D-4}}{\zeta\left( 5-D \right)}
\,\delta(x-nM^2)
\nonumber\\ 
&& 
\hspace{-1.25cm}
+ \left( \frac{M^2}{\nu^2} \right)^{\frac{D}{2}-2}
\Gamma \left(2 - \frac{D}{2} \right)  \frac{e^{\gamma_E(4-D)/2}}{D-4}
\nonumber\\
&& \times 
\bigg\{ 4 \xi_{12}^I 
+ (D-4)\big[\xi_{11}^I + 2 \gamma_E \xi_{12}^I\big]
\nonumber\\
&& 
- \frac{(D-4)^2}{2} \bigg[  \xi_{10}^I + \xi_{11}^I \gamma_E 
- 2 \xi_{12}^I \Big( \frac{\pi^2}{6} + \gamma_1  \Big) \bigg] \bigg\}
\nonumber\\ 
&& \left. \times 
\sum_{n\ge 1} M^2 \frac{n^{\frac{D}{2} - 2}}{\zeta\left( 3 - \frac{D}{2} \right) }
\,\delta(x-nM^2)
\right\}
.
\eea
It is a straightforward exercise, making use of the techniques described in the 
second part of Section \ref{matching_LO}, to show that this spectral density leads
to the desired properties. For instance, after minimal subtraction of the simple 
and double poles at $D=4$, the renormalized dispersive integral reads
\bea
&&
\hspace{-1.25cm}
\int dx \, \frac{\rho^{\rm res;1}_{K\pi}(x ; D)}{x+wM^2}
\nonumber\\
&&
\hspace{-1.25cm}
= - \, 16 \pi^2 M_K^2 \left( - \frac{G_{\rm F}}{\sqrt{2}} V_{us} V_{ud} \right)
\alpha_s (\nu) \sum_{I=1}^6 C_I (\nu)  \Bigg\{ \xi_{12}^I  {\widetilde\psi} (w)
\nonumber\\
&&
\hspace{-1.25cm}
+ \left[ 2  \xi_{12}^I \ln \frac{M^2}{\nu^2} - \xi_{11}^I
\right]   \psi(1+w) 
+ \xi^I_{10} 
- 2 \left( \frac{\pi^2}{6} + \gamma_1  \right) \xi_{12}^I
\nonumber\\
&&
\hspace{-1.25cm}
+
\left[
 \xi^I_{12} \ln \frac{M^2}{\nu^2}   - \xi^I_{11} \right]
\ln \frac{M^2}{\nu^2} \Bigg\}
+
{\cal O}(D-4)
,
\eea 
with
\bea
\hspace{-0.5cm}
{\widetilde\psi} (w) \!\!\!&=&\!\!\! 2 M^2 w \int \frac{dx}{x} \, \frac{1}{x + M^2 w} \sum_{n\ge 1} M^2 (\ln n) \delta (x - nM^2)
\nonumber\\
\!\!\!&=&\!\!\! 2 w \sum_{n\ge 1} \frac{\ln n}{n(n+w)}
.
\eea
We now need to establish some properties of the function ${\widetilde\psi} (w)$,
which is defined by an absolutely convergent sum as long as $w$ is not equal to 
a negative integer. Thus, ${\widetilde\psi} (0)=0$ and
\be
{\widetilde\psi}^\prime (0) = 2 \sum_{n\ge 1} \frac{\ln n}{n^2} = - 2 \zeta^\prime (2)
.
\ee
Next, using the relation \rf{mellin}
%\be
%\frac{1}{1+\frac{w}{n}} = \int_{c_1-i\infty}^{c_1+i\infty} \frac{du}{2 \pi i} \left(\frac{w}{n}\right)^{-u} \Gamma(u) \Gamma(1-u)
%,
%\lbl{mellin}
%\ee
%where $0< c_1 <1$, 
and
\be
\sum_{n\ge 1} n^{u-2} \ln n = - \zeta^\prime (2-u)
,
\ee
one obtains the following Mellin representation of the function ${\widetilde\psi} (w)$
\be
{\widetilde\psi} (w) = -2 w \! \int_{c_1-i\infty}^{c_1+i\infty} \!\frac{du}{2 \pi i} w^{-u} \Gamma(u) \Gamma(1-u) 
\times \zeta^\prime (2-u)  .
\lbl{Mellin_psi-tilde}
\ee
The first singularity on the left of the fundamental strip $0<c_1<1$ lies at $u=0$. It consists of a simple pole coming from 
the factor $\Gamma(u)$ of the integrand. Therefore, the Converse Mapping Theorem allows to state that 
${\widetilde\psi} (w) = -2w\zeta^\prime (2) + \cdots$ as $w\to 0$, a property already established before. 
The first singularity one encounters on the right of the fundamental strip consists of a triple pole and a
simple pole, both located at $u=+1$,
\be
\Gamma(u) \Gamma(1-u) \zeta^\prime (2-u) \underset{u \rightarrow +1}{\sim}
\frac{1}{(u-1)^3} + \frac{\frac{\pi^2}{6} + \gamma_1}{u-1} +  \gamma_2 + {\cal O}(u-1)
.
\ee
The Converse Mapping Theorem then allows to conclude that
\be
%\sum_{n\ge 1} \frac{\ln n}{n^2} = - \zeta^\prime (2),\qquad
{\widetilde\psi} (w) \underset{w \rightarrow +\infty}{\sim}
\ln^2 w + 2  \left( \frac{\pi^2}{6} + \gamma_1  \right) + \cdots
,
\ee
which is precisely what is required. The function ${\widetilde\psi} (w)$
does not seem to be related in an obvious way to the standard sets of functions that have been studied in
the mathematical literature, see e.g. \cite{DLMF} and references therein. It is thus
interesting to notice that the Mellin representation \rf{Mellin_psi-tilde} of ${\widetilde\psi} (w)$, 
which holds for ${\rm Re} w >0$, can be recast into [the integral is understood as its Cauchy principal 
value for $w$ real and positive]
\bea
&& \hspace{-1.25cm}
{\widetilde\psi} (w) = 
2 \int_0^\infty \frac{dx}{x} \frac{w}{x-w} \, [\psi(1+x) + \gamma_E]
\nonumber\\
&& \hspace{-0.2cm}
+ 2  [\psi(1+w) + \gamma_E]\ln w
.
\eea
This representation allows to extend the function ${\widetilde\psi} (w)$ to negative 
values of ${\rm Re} w$. The integral is then regular, while the second term
reproduces the poles of ${\widetilde\psi} (w)$ when $w$ equals a strictly negative integer. 
The determination of $\ln w$ is given by $w\to w-i\epsilon$, with $\epsilon>0$ and infinitesimal, 
in agreement with the prescription $s \to s+i\epsilon$.
This representation also proves quite useful for the numerical evaluation of ${\widetilde\psi} (w)$.

\begin{figure}[t]
%\vspace{-2.5cm}
~\includegraphics[width=8.0cm]{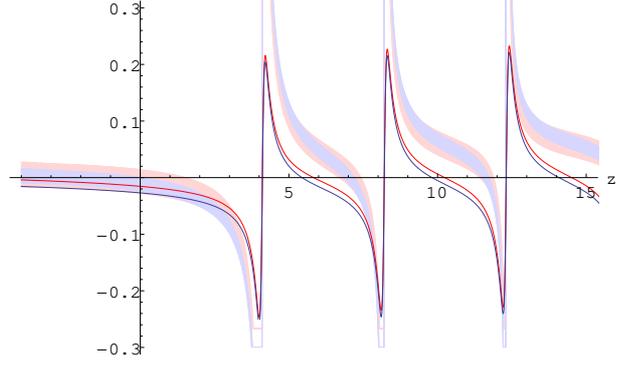} 
\caption{The function ${\cal W}_{K\pi}^{\rm res} (z ; \nu)$ defined in Eq. \rf{Wres_final}
as a function of $z$ for $\nu=2~{\rm GeV}$ and $M=1~{\rm Gev}$. The light-red (purple)
band corresponds to the NDR (HV) scheme. The red (blue) line {corresponds} to the truncation 
of ${\cal W}_{K\pi}^{\rm res} (z ; \nu)$ to lowest order in the NDR (HV) scheme.
For the ease of visualization, the location of the poles at $z=nM^2/M_K^2$ has been slightly 
shifted off the real-$z$ axis.} \label{plot_calW}
\end{figure}

\section{Discussion}\label{discussion}
\setcounter{equation}{0}

\begin{figure*}[t]
\includegraphics[width=8.26cm]{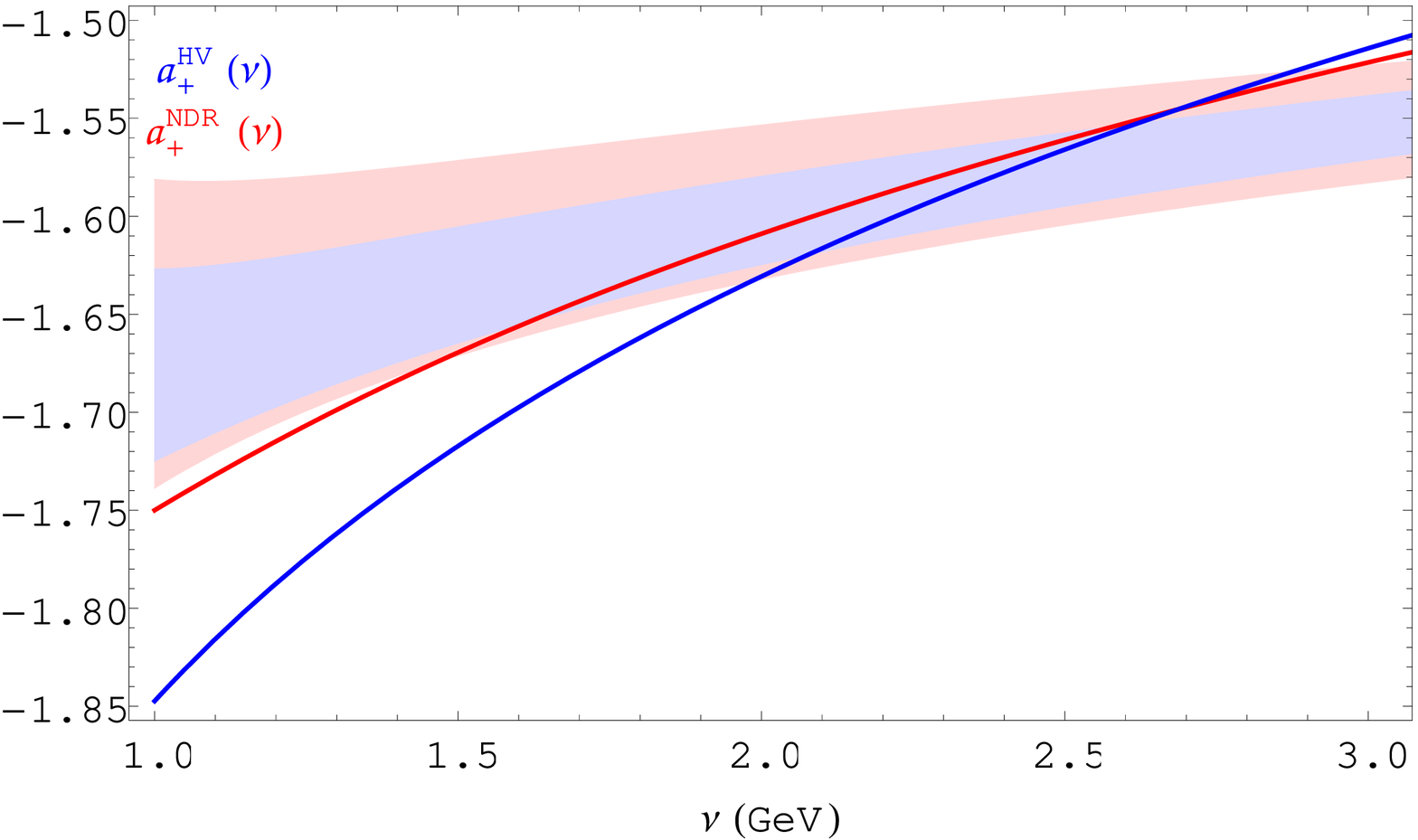} ~~~ \includegraphics[width=8.2cm]{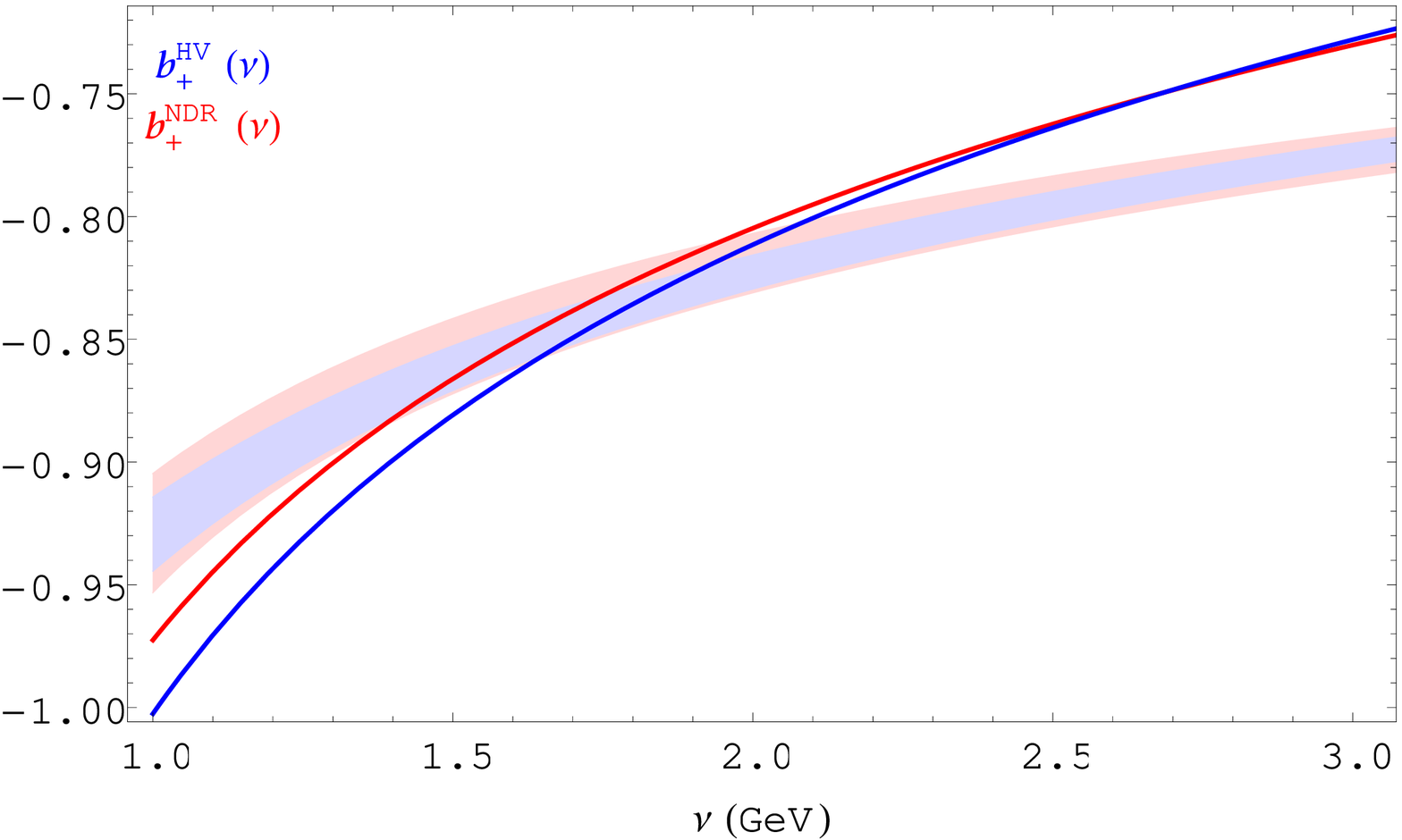}
\caption{The variation of $a_+$ (left plot) and of $b_+$ (right plot) as a 
function of the short-distance renormalization scale $\nu$. The scale dependence at lowest order
corresponds to the red (blue) lines in the NDR (HV) scheme. The scale dependence after
inclusion of ${\cal O}(\alpha_s)$ corrections corresponds to the light-red (purple)
bands in the NDR (HV) scheme.} \label{plots_scale}
\end{figure*}

We have shown that it is possible to construct a function $W_{K\pi}^{\rm res} (z ; \nu)$
through an infinite sum of zero-width resonances with a Regge-type spectrum, such that its short-distance
behaviour matches, at order ${\cal O}(\alpha_s)$, the one obtained from the leading term in the OPE. After renormalization
in the ${\overline{\rm MS}}$ scheme of dimensional regularisation, and up to the order ${\cal O} (\alpha_s)$,
this function reads [$w \equiv -z M_K^2/M^2$]
\bea
&& \hspace{-1.35cm}
W_{K\pi}^{\rm res} (z ; \nu) = 
- 16 \pi^2 M_K^2 \! \left( \! - \frac{G_{\rm F}}{\sqrt{2}} V_{us} V_{ud} \right)
\!\frac{C_{K\pi} f_+^{K\pi} (z M_K^2)}{4\pi}
\nonumber\\
&& \hspace{0.1cm}
\times  \sum_{I=1}^6 C_I (\nu)\Bigg\{
\xi_{00}^I 
+ \alpha_s (\nu) \xi^I_{10} 
%- 2 \alpha_s (\nu) \left( \frac{\pi^2}{6} + \gamma_1  \!\right) \!\xi_{12}^I
\nonumber\\
&& \hspace{0.1cm}
+ 
\ln \frac{\nu^2}{M^2} \left[ \xi_{01}^I + \alpha_s (\nu) 
\left( \xi^I_{11}
+ \xi^I_{12} \ln \frac{\nu^2}{M^2} \right) \right]
\nonumber\\
&& \hspace{0.1cm} 
- \psi(1+w) \left[ \xi_{01}^I + \alpha_s (\nu) \left( \xi_{11}^I + 2 \xi_{12}^I \ln \frac{\nu^2}{M^2} \right) \right]
\nonumber\\
&& \hspace{0.1cm} 
+ \alpha_s (\nu) \, \xi_{12}^I \left( {\widetilde\psi} (w) - \frac{\pi^2}{3} - 2 \gamma_1 \right)
\!\Bigg\}
\nonumber\\
&&
\equiv
- 16 \pi^2 M_K^2 \! \left( \! - \frac{G_{\rm F}}{\sqrt{2}} V_{us} V_{ud} \right)
\!\frac{C_{K\pi} f_+^{K\pi} (z M_K^2)}{4\pi}
\nonumber\\
&& \hspace{0.1cm}
\times 
{\cal W}_{K\pi}^{\rm res} (z ; \nu)
.
\lbl{Wres_final}
\eea
This expression involves two scales. The first one, $M$, is the scale of the
lowest resonance in the spectrum besides the $\rho(770)$, which is already taken
into account by the contribution $W_{K\pi}^{\pi\pi} (z)$ in Eq. \rf{model}, see 
Ref. \cite{DAmbrosio:2018ytt}. The first resonance appearing in $W_{K\pi}^{\rm res} (z ; \nu)$
is thus the $K^*(892)$ or the $\phi(1020)$, hence $M\sim 1~{\rm GeV}$. The second
scale in Eq. \rf{Wres_final} is the renormalization scale $\nu$. It represents
the onset of the perturbative regime of QCD. In practice, the description of the spectrum
in terms of well-identified resonances extends to a few radial excitations of the resonances
mentioned just above, for instance $\rho(1450)$, $\rho(1570)$, $\phi(1680)$..., before
they merge into the continuum. This means $\nu\sim 2~{\rm GeV}$.

The function ${\cal W}_{K\pi}^{\rm res} (z ; \nu)$ is shown in Fig. \ref{plot_calW}.
For positive values of $z$ it displays the expected infinite series of equally-spaced poles, 
whereas for negative values of $z$ the asymptotic regime sets in rapidly. 
In order to draw these plots, some knowledge of the coefficients $\xi_{10}^I$,
which are not fixed by the renormalization-group constraint \rf{RG_equ}, is needed.
For the sake of illustration, we have estimated these coefficients to vary in the range
$-\xi_{11}^I \le \xi_{10}^I \le + \xi_{11}^I$, taking the relation \rf{xi_LO} between
the lowest-order coefficients $\xi_{00}^I$ and $\xi_{01}^I$ as a guide. The same choice
also applies to Fig. \ref{plots_scale}.

In the remainder of the Letter, we wish to address in turn two issues that we think 
deserve to be given some consideration, namely: i) the size of the residual dependence
with respect to the short-distance scale $\nu$ and ii) some features and properties of 
the resonance model that we have constructed, as well as possible improvements.

\subsection{Residual scale dependence}

By construction, the expression of $W_{K\pi}^{\rm res} (z ; \nu)$ displayed
in Eq. \rf{Wres_final} provides a form factor that is independent of the 
renormalization scale $\nu$ at order ${\cal O}(\alpha_s)$,
\be
\nu \frac{d}{d \nu} \left[ W_{K\pi}^{\rm res} (z ; \nu) + W_{K\pi}^{\rm SD} (z ; \nu) \right]
= {\cal O}(\alpha_s^2)
.
\ee
We expect the residual scale dependence to be weaker { than} the one that 
results from the matching at lowest order only. We illustrate these
changes in the case of the two constants $a_+$ and $b_+$ corresponding
to the values of the form factor $W_{K\pi} (z)$, for $(K,\pi) = (K^+ , \pi^+)$, 
and of its derivative at $z=0$, respectively. 
For this purpose, we add to the sum of $W_{K\pi}^{\rm res}$ and $W_{K\pi}^{\rm SD}$
the contribution $W_{K\pi}^{\pi\pi}$ of the two-pion intermediate 
state evaluated in Ref. \cite{DAmbrosio:2018ytt}, see Eq. \rf{model} above.
This means $a_+\vert_{\pi\pi}=-1.58$ and $b_+\vert_{\pi\pi}=-0.76$.
The improvement when going from lowest order
to next-to-leading order can be appreciated from the plots shown in Fig. \ref{plots_scale}
[details on the numerical aspects can be found in the appendix]. We notice that
indeed both the scale dependence and the scheme dependence become less pronounced when order 
${\cal O}(\alpha_s)$ corrections are included. Our ignorance of the coefficients $\xi_{10}^I$
induces, under the conditions stated above, an uncertainty that amounts, at $\nu=2~{\rm GeV}$, 
to about $5\%$ in both $a_+$ and $b_+$. We also observe that the corrections
coming from the resonance and short-distance parts are quite small as compared to the
contribution from the two-pion state.

This improvement in the control over the scheme and scale {dependences} allows 
us to refine somewhat our evaluation of the coefficients $a_+$ and $b_+$ made in
Ref. \cite{DAmbrosio:2018ytt}. We obtain 
\be
a_+ = - 1.59(8) , \quad b_+ = -0.82(6).
\ee
These values are still at variance with the experimental determinations
$a_+^{\rm exp} =-0.593(11)$ and $b_+^{\rm exp}=-0.675(43)$, see Ref. \cite{DAmbrosio:2018ytt}
for a detailed discussion, especially as far as $a_+$ is concerned. On the other 
hand, as discussed in Ref. \cite{DAmbrosio:2018ytt}, in order to reach a
definite conclusion, the contributions 
$a_+\vert_{\pi\pi}$ and $b_+\vert_{\pi\pi}$ need to be evaluated within 
a tighter framework than the one adopted there,
and other exclusive contributions, in particular from two-kaon states,
should eventually be accounted for explicitly, see also the discussion below.
Work in this direction is on its way \cite{in_progress}.

\subsection{Properties and features of the resonance model}

The resonance model that we have constructed in Sections \ref{matching_LO} and \ref{matching_NLO}
is by far not unique. Discussing the arbitrariness of the construction in full generality represents
a formidable, if not impossible, task, given the fact that the corresponding dispersive integrals
are only constrained to reproduce the divergences of perturbative QCD at $z=0$ and the leading
asymptotic behaviour from the OPE at $z\to -\infty$. The determination of sub-leading terms in the OPE 
might be a way to constrain the resonance model further. We will, however, not address this possibility 
here, but rather state a few remarks concerning the model that we have constructed above.

Some of the arbitrariness of the construction is embodied in the functions $A_0(D)$ and $f(D)$ in Section \ref{matching_LO}, 
or $A_{1a}(D)$, $A_{1b}(D)$ ${\tilde f}(D)$ and $g(D)$ in Section \ref{matching_NLO}. These functions do 
not appear anymore in the final expression
\rf{Wres_final}, partly because they are absorbed in the matching to the coefficients $\xi_{rp}^I$, partly
because they contribute only to the terms of order ${\cal O}(D-4)$ in the renormalized dispersive integral.

These features certainly do not exhaust all the arbitrariness of the model. While the contribution of the 
resonances with higher masses will be constrained by the short-distance behaviour, one might expect that
the {description} of the lower-lying resonances like $K^*(892)$ or $\phi(1020)$ [recall that the contribution
from $\rho(770)$ is already taken care of by $W_{K\pi}^{\pi\pi}$] provided by this model may be less realistic
from a phenomenological point of view.
One way to circumvent this possible drawback would be to consider additional intermediate states in a more
explicit way, i.e. by extending the decomposition in Eq. \rf{model} to, for instance \cite{DAmbrosio:2018ytt}
\be
W_{K\pi}^{\rm LD} (z) = W_{K\pi}^{\pi\pi} (z) + W_{K\pi}^{K\pi} (z) + W_{K\pi}^{K{\bar K}} (z) + W_{K\pi}^{\rm res} (z ; \nu) 
.
\ee
We will, however, not pursue this matter in the present Letter, and leave the discussion of such an extension and of the corresponding 
multi-channel analysis it requires for future work.

\appendix

\section*{Appendix}
\setcounter{section}{1}
\def\theequation{\Alph{section}.\arabic{equation}}

In this appendix, we gather some information and give the input values used in order to
produce the plots shown in Fig. \ref{plots_scale}. The running of the Wilson coefficients
is given in Eqs. \rf{scale_C7V} and \rf{scale_CI}
using the anomalous dimensions $\gamma_{J,7V}$ given in \cite{Buras:1994qa} and $\gamma_{I,\,J}$ given in
\cite{Buras:1992tc,Ciuchini:1993vr}. We have, however, restricted
ourselves to the contributions from the two current-current operators $C_\pm = C_2 \pm C_1$,
neglecting those from the QCD penguin operators $C_I$, $I=3,4,5,6$. The Wilson coefficients
at next-to-leading order are then given by
\bea
&&
\hspace{-1.25cm}
\frac{C_+ (\nu)}{C_+ (\nu_0)} = 
\left[ 1 + S_{++} \left( \frac{\alpha_s (\nu)}{4 \pi} - \frac{\alpha_s (\nu_0)}{4 \pi} \right) \right]
\left( \frac{\alpha_s (\nu)}{\alpha_s (\nu_0)} \right)^{-2/9} 
,
\nonumber\\
\\
&&
\hspace{-1.25cm}
\frac{C_- (\nu)}{C_- (\nu_0)} = \left[ 1 + S_{--} \left( \frac{\alpha_s (\nu)}{4 \pi} - \frac{\alpha_s (\nu_0)}{4 \pi} \right) \right]
\left( \frac{\alpha_s (\nu)}{\alpha_s (\nu_0)} \right)^{+4/9}
,
\nonumber
\eea
and
\bea
&&
\hspace{-1.25cm}
\frac{C_{7V} (\nu)}{\alpha} - \frac{C_{7V} (\nu_0)}{\alpha}
\\
&&
\hspace{-1.0cm}
=  \frac{16}{99} \frac{C_+ (\nu_0)}{\alpha_s (\nu_0)}
\Bigg\{ 1 -
\bigg[ 1 - S_{++} \left( \frac{\alpha_s (\nu_0)}{4 \pi} - \frac{\alpha_s (\nu)}{4 \pi} \right) \bigg]
\nonumber\\
&&
\hspace{-0.77cm}
\times
\left( \frac{\alpha_s (\nu)}{\alpha_s (\nu_0)} \right)^{-11/9}
%\nonumber\\
%&& \hspace{2.75cm}
\!\!\! - \,S_{7+} \, \frac{\alpha_s (\nu_0)}{4 \pi}\bigg[ \left( \frac{\alpha_s (\nu)}{\alpha_s (\nu_0)} \right)^{-2/9} - 1 \bigg]
\Bigg\}
\nonumber\\
&&
\hspace{-1.0cm}
 -\, \frac{8}{45} \frac{C_- (\nu_0)}{\alpha_s (\nu_0)}
\Bigg\{ 1 - 
\bigg[ 1 - S_{--} \left( \frac{\alpha_s (\nu_0)}{4 \pi} - \frac{\alpha_s (\nu)}{4 \pi} \right) \bigg]
\nonumber\\
&&
\hspace{-0.77cm}
\times
\left( \frac{\alpha_s (\nu)}{\alpha_s (\nu_0)} \right)^{-5/9}
%\nonumber\\
%&& \hspace{2.75cm}
\!\!\! - \,S_{7-} \, \frac{\alpha_s (\nu_0)}{4 \pi}\bigg[ \left( \frac{\alpha_s (\nu)}{\alpha_s (\nu_0)} \right)^{+4/9} - 1 \bigg]
\Bigg\}
,
\nonumber
\eea
where
\be
S_{++} = \frac{307}{162} , \ S_{--} = - \frac{181}{81} , \ S_{7+} = - \frac{2071}{108} , \ S_{7-} = \frac{443}{108}
\ee
in the NDR scheme, and
\be
S_{++} = -\frac{773}{162} , \ S_{--} = - \frac{397}{81} , \ S_{7+} = - \frac{3331}{108} , \ S_{7-} = \frac{1991}{108}
\ee
in the HV scheme. These expressions for $C_\pm(\nu)$ and for $C_{7V}(\nu)$
hold at next-to-leading order,
and their truncation to the lowest order is obtained upon taking all the
coefficients $S_{ij}$ equal to zero. The input values we have used are 
\cite{Buras:1994qa}
\be
C_+(\nu_0) = 0.688,\ C_-(\nu_0) = 2.108,\ C_{7V} (\nu_0)/\alpha = -0.026
\ee
at lowest-order. At next-to-leading order they become
\be
C_+(\nu_0) = 0.771,\ C_-(\nu_0) = 1.737,\ C_{7V} (\nu_0)/\alpha = -0.037
\ee
in the NDR scheme and
\be
C_+(\nu_0) = 0.735,\ C_-(\nu_0) = 1.937,\ C_{7V} (\nu_0)/\alpha = 0.000
\ee
in the HV scheme. In all three cases these values hold for $\nu_0 = 1~{\rm GeV}$.
The running of $\alpha_s$ is given by
\be
\alpha_s (\nu) = \frac{2\pi}{9 \ln(\nu/\Lambda)} \left[
1 - \frac{32}{81} \frac{\ln\left(\ln\nu^2/\Lambda^2\right)}{\ln(\nu/\Lambda)}
\right]
\ee
at next-to-leading order, and with the obvious truncation at lowest order.
The QCD scale for three active flavours is taken as $\Lambda = 332~{\rm MeV}$.
Finally, the mass of the lowest resonance is set at $M=1~{\rm GeV}$.

 %%%%%%%%%%%%%%%%%%%%%%%
 \section*{Acknowledgements}
 %%%%%%%%%%%%%%%%%%%%%%%
One of us (M.K.) wishes to thank the INFN-Sezione di Napoli and the Universit\'a di Napoli Federico II for 
their warm hospitality.
 G.D.   was supported in part by MIUR under Project No. 2015P5SBHT and by the INFN research initiative ENP. The work of M.K. has received partial support from the OCEVU Labex (ANR-11-LABX-0060) 
and the A*MIDEX project (ANR-11-IDEX-0001-02) funded by the “Investissements d'Avenir” French government 
program managed by the ANR.
 The Feynman diagrams displayed in Fig. \ref{OPE_QCD_fig} have been drawn using Jaxodraw 2.1-0
\cite{Binosi:2003yf,Binosi:2008ig}.

% \vspace{0.1cm}

\end{document}